\documentclass[preprint2]{aastex}
\shorttitle{Jupiter Deflection}
\shortauthors{Fomalont \& Kopeikin}
\begin{document}

\title{The Measurement of the Light Deflection from
Jupiter: \\Experimental Results}

\author{E.~B.~Fomalont}
\affil{National Radio Astronomy Observatory, Charlottesville, VA 22903}
\email{efomalon@nrao.edu}

\author{S.~M.~Kopeikin}
\affil{Dept. of Physics and Astronomy, University of Missouri-Columbia,
       Columbia, MO 65211}
\email{kopeikins@missouri.edu}

\begin{abstract}
    We have determined the relativistic light deflection of the quasar
J0842+1835 as Jupiter passed within $3.7'$ on 2002 September 8, by
measuring the time delay using the VLBA and Effelsberg radio
telescopes at 8.4 GHz.  At closest approach, General Relativity (GR)
predicts a radial (static) deflection of $1190~\mu$arcsec, and a
tangential (retarded) deflection in the direction of Jupiter's motion
of $51~\mu$arcsec.  Our experiment achieved an rms position error of
$\le 10~\mu$arcsec, and measured this retarded deflection to be
$0.98\pm 0.19$ (rms error) times that predicted by GR.  The increased
positional accuracy for this VLBI phase referencing experiment was
achieved by using two calibrator sources.  Comments on the
interpretation of this experiment, are given.
\end{abstract}


\section{The Gravitational Deflection of Radio Waves by Jupiter}

    Einstein solved the equation of light propagation in the field of
a static body, and predicted the deflection of light at the limb of
the Sun of $1.75''$ \citep{ein16}, and this relativistic deflection
was measured for the first time in 1919 \citep{dys20}.  The deflection
prediction by Einstein is equivalent to measuring the time delay of
light traveling in a static gravitation field \citep{sha64}, and was
confirmed in a series of experiments, the most precise of which was
made with Very Long Baseline Interferometry (VLBI) \citep{leb95,
eub97}.  \cite{kop01,kop03} generalized the problem of light
propagation in the gravitational field of arbitrary moving bodies, and
showed that the Lorentz-invariant expression for the relativistic time
delay of light (and light deflection), to all orders in $v/c$, depends
on the retarded positions of the moving bodies as defined by the
retarded Li\'enard-Wiechert solution of the Einstein gravitational field
equations.  For Jupiter, as an example, the Lorentz-invariant
relativistic time delay between two telescopes (\#1 and \#2), from
a source of light at infinity (quasar), is given
by the equation \citep{kop01}

\begin{eqnarray}
\Delta= {2GM_J\over c^3}\biggl[1+{{\it\bf K}{\bf\cdot}{\it\bf v}_J(s_1)\over c}\biggr] \nonumber \\
\ln\frac{r_{1J}(s_1)+{\it\bf K}{\bf\cdot}{\it\bf r}_{1J}(s_1)}{r_{2J}(s_2)+{\it\bf K}{\bf\cdot}{\it\bf r}_{2J}(s_2)}\,
\end{eqnarray}

\noindent where $G$ is the gravitational constant, $c$ is the speed of
light, $M_J$ is the mass of Jupiter, ${\it\bf K}$ is the unit vector
from the barycenter of the solar system to the quasar, and ${\it\bf
v}_J(s_1)$ is the coordinate velocity of Jupiter taken at the
retarded time $s_1$.  The retarded times $s_1$ and $s_2$ are
calculated according to the {\it gravity null-cone} equations, as defined
by the Li\'enard-Wiechert retarded solution of the Einstein gravity
field equations which contain a constant in the wave operator that we
denote as $c_g$ \citep{kop03,wil03}. The retarded times are

\begin{eqnarray}
s_1&=&t_1-{1\over c_g}|{\it\bf x}_1(t_1)-{\it\bf x}_J(s_1)|\;,\\
s_2&=&t_2-{1\over c_g}|{\it\bf x}_2(t_2)-{\it\bf x}_J(s_2)|\;,
\end{eqnarray}

\noindent where $t_1$, $t_2$ are times of arrival of the radio signal
from the quasar to the first and second VLBI stations respectively,
having coordinates ${\bf x}_1(t_1)$ and ${\bf x}_2(t_2)$, and ${\bf
x}_J(s)$ is coordinate of Jupiter taken at the retarded time.  The
distances between each telescope and Jupiter are given by
$r_{1J}=|{\it\bf r}_{1J}|$, $r_{2J}=|{\it\bf r}_{2J}|$, ${\it\bf
r}_{1J}(s_1)={\it\bf x}_{1}(t_1)-{\it\bf x}_{J}(s_1)$ and ${\it\bf
r}_{2J}(s_2)={\it\bf x}_{2}(t_2)-{\it\bf x}_{J}(s_2)\;$, all of which
depend implicitly on the value of $c_g$.  GR predicts that the
constant, $c_g$, equals the speed of light $c$.  If $c_g\ne c$, then
Eqs.\ (1) to (3) would not be invariant with respect to the Lorentz
transformation.  Hence, the experiment measures the numerical value of
$c_g$ as a test of the Lorentz invariance of the Einstein equations,
and is an indirect measurement of the speed of propagation of
gravity\footnote{The radio wave from the quasar does not propagate
along the gravity null cones (Eqs.\ (2) and (3)) and, therefore, $c_g$
can not be physically identified with the speed of light \citep{kop01,
kop03}, as was done, for example by \cite{asa02}}.  
Standard data analysis of VLBI observations uses the barycentric time
$t$ and the barycentric space coordinates ${\bf x}=(x,y,z)$.  Hence, all
quantities and variables in Eqs.\ (1)--(3), and hereafter, are
expressed specifically in these coordinates. For example, ${\bf
v}_J = d{\bf x}/dt$ is the orbital velocity of Jupiter with respect to
the barycenter of the solar system, and vector ${\bf K}$ defines
the direction to the quasar as measured by an observer at rest with
respect to the barycenter (see \cite{kop01} and \cite{kop02} for
more detail).

Relativistic time delay of light in the gravitational field of moving
Jupiter was measured in a VLBI experiment on September 8, 2002 when
Jupiter passed within $3.7'$ ($\sim 14$ jovian radii) of the bright
radio quasar J0842+1835.  If Jupiter is at a small angular distance
$\theta$ from the quasar, Eq.\ (1) can be expanded in a Taylor series
around the time of arrival{\footnote{In the general case when angle
$\theta$ is not small the accurate retarded Eqs. (1)--(3) must be
used.}. Then, the linearized relativistic time delay, $\triangle
(\tau_{l,m})$ between two radio telescopes from a VLBI network having
numbers $l$ and $m$ and separated by a distance ${\bf B}_{l,m}$, can
be expressed to sufficient accuracy as
\begin {eqnarray}
\triangle (\tau_{l,m}) = -\frac{4GM_J}{c^3r_{mJ}}~~~~~~~~~~~~ \nonumber \\
\bigg[\left(1-\frac{2{\bf n}{\bf\cdot}{\bf 
v}_{TJ}}{c_g\theta}\right)\frac{{\bf n\cdot B}_{l,m}} {\theta} +
\frac{{\bf B}_{l,m}{\bf\cdot v}_{TJ}}{c_g\theta^2~}\bigg] 
\end {eqnarray}

\noindent where ${\bf v}_{TJ}={\bf K}\times({\bf v}_J\times {\bf K})$
is Jupiter's velocity in the plane of the sky, ${\bf n}$ is the unit
vector from Jupiter to the radio source in the plane of the sky, and
$r_{mJ}$ is the distances between Jupiter to the $m$th telescope.  The
angle $\theta$ is defined by the relationship (\cite{kop01}, see Eq.\
(9)), $\hbox{cos}~\theta=-{\bf K}\cdot{\bf p}$, where ${\bf p}=({\bf
x}_1-{\bf x}_J(t_1))/|{\bf x}_1-{\bf x}_J(t_1)|$, and is calculated
from the accurate JPL ephemeris \citep{sta00}.  The first term in the
large brackets is the radial deflection term\footnote{This equation
treats more precisely the radial deflection than the corresponding
equation from \citep{kop01}.}, and the second term in brackets is the
tangential deflection which is in the direction of the velocity of
Jupiter with respect to the barycenter of the solar system, ${\bf
v}_{TJ}$, in the plane of the sky.  The velocity-dependent terms in
Eq.\ (4) result from the Taylor expansion around the time of arrival
of $t_1$ of the retarded arguments in Eq.\ (1).  In the analysis of
the experimental data, we used Eq.\ (1)-(3), rather than the
approximate form in Eq.\ (4).  This misunderstanding of the analysis
process has led to the erroneous claim \citep{sam03} that no $v/c$
terms are observed with this deflection experiment (see \cite{kop03}
for more detail).

    The constant $c_g$, shown explicitly in Eq.\ (4), is associated in
our model of the experiment with a speed of gravity ( scalar and
vector modes) propagating along the gravity null-cone Eqs.\ (2) and
(3) in accordance with the physical interpretation of the
Li\'enard-Wiechert solutions of Einstein's equations
\citep{kop01,kop03}\footnote{The tensor modes (free gravitational
waves) also propagate along the gravity null cones but they can not be
detected in the near zone of the solar system.  Hence, the experiment
can not be viewed as a detection of gravitational waves which are too
faint to be seen with present VLBI accuracy \citep{kop99b}.}.  We
introduce a fitting parameter $\delta=c/c_g-1$ to measure the
difference between the two speeds. For GR, $\delta=0$. Other
theoretical models of this experiment \citep{wil03,asa02} predict the
same time delay given in Eqs. (1) and (4), but only to the first order
of $v/c$. However, these different formulations lead to a different
interpretation of $c_g$. See \S 4b for further discussion.

    The order of magnitude of the deflection prediction on September 8
for the 6000-km telescope separation is a delay of 115 psec
(deflection of $1190~\mu$arcsec) for the radial term, and a delay of
4.8 psec (deflection of $51~\mu$arcsec) for the retarded term at the
point of closest approach.  Although $v_J/c \approx 4\times 10^{-5}$,
the additional factor of $1/\theta$ amplifies the retarded term so
that it is 4\% of the radial term\footnote{The amplification is due to
the time lag between present and retarded positions of Jupiter
(Kopeikin 2001).}.  A previous close passage occurred in 1988 March 21
\citep{tre91} and the radial deflection term was measured to an
accuracy $\approx 15$\% in accordance with GR.  With improvements over
the years in VLBI techniques, sub-milliarcsecond positional accuracy
were now attainable, and measurement of the retarded term was feasible.

\section{The Experimental Strategy}

The VLBI experiment to measure the deflection of light of the quasar
J0842+1835 by Jupiter was conducted during five days, centered on
September 8, 2002.  The radio array consisted of eleven telescopes:
ten 25-m diameter telescopes of the Very Long Baseline Array (VLBA),
plus the 100-m diameter telescope at Effelsberg, Germany.  The
observing frequency was 8.45 GHz, and ten hours were used for
observations on each of 2002 September 4, 7, 8, 9, 12.

    An array measures the difference in arrival time, the delay, for a
quasar signal to reach each of the telescopes.  Using a widely spaced
array, extremely high positional accuracy can be obtained.  However,
this delay is contaminated by many effects, both internal to and
external to each telescope system.  To remove these effects, radio
astronomers use the technique of {\it phase referencing} (alternating
observations of two sources) whereby observations of one source (the
calibrator) are used to determine the delay errors associated with the
target source \citep{bea95}.  If the switching time and source
separation between calibrator and target are less than the
temporal/spatial variation scale sizes, the relative position between
the two sources can be accurately measured.  However, the propagation
delay along the quasar paths through the ionosphere and troposphere
may be so variable in time and angle, that even fast switching will
not completely remove these propagation changes, and relative position
accuracies $\le 30~\mu$arcsec are difficult to obtain even with
VLBI techniques.

     After several sets of test observations in early 2002
\citep{fom02}, we decided to sequentially observe J0842+1835 with two
known nearby calibrators: J0839+1802, separated $0.82^\circ$ in
position angle $-132^\circ$; and J0854+2006 (=OJ287), separated by
$3.36^\circ$ in position angle $63^\circ$.  The sources are
identified as quasars with high redshift, and they lie
on a nearly straight line, which was exploited in the calibration
procedure.  The observing sequence was identical for the 5 observing
days: We cycled observations through the three sources in 5.5 min,
with about 100 such cycles per day, a reasonable compromise between
removing the temporal effects with obtaining good accuracy of the
measured delay.  This observational technique was an improved variant
of that used for the solar bending experiments in 1974 and 1975
\citep{fom76}.  The jovian magnetosphere also produced a radial
deflection of the radio waves (directed inward to Jupiter) and this
anomalous bending was of concern.  However, calculations suggested
that this bending would be significantly less than the retarded
deflection term, and special observing techniques (observing at two
frequencies simultaneously) would have reduced the sensitivity of the
experiment to the gravitational bending.  The magnetosphere bending is
discussed in more detail below.

\section{The Data Reduction}

\subsection {The Radio Interferometer Response}

     A radio interferometer measures the complex spatial coherence
function of the electromagnetic radiation field
$C_{l,m}~e^{i2\pi\psi_{l,m}}$ at a frequency $\nu$ between two
telescopes denoted by $l$ and $m$.  If the intercepted electromagnetic
radiation field is dominated by a small-diameter radio source, then
the response of the interferometer, denoted as the complex visibility
function $V_{l,m}~e^{i2\pi\phi_{l,m}}$, is
\begin{eqnarray}
{V}_{l,m}~e^{i2\pi\phi_{l,m}} = C_{l,m}~e^{(i2\pi\psi_{l,m})} \nonumber \\
\hat{G}_l\hat{G}^*_m~e^{\frac{i2\pi\nu}{c}(\triangle\tau_l-\triangle\tau_m)}
\end{eqnarray}
\noindent
The complex gains of the telescopes, $\hat{G}_l$, contain the
amplification and phase shifts that are introduced by each telescope
system (* denotes complex conjugate), and the residual time delay
(observed - model) of the signal from the radio source to each
telescope is denoted by $\triangle\tau_l$.  For an array of $N$
telescopes, the visibility function is sampled simultaneously with
$N(N-1)/2$ interferometers, 55 pairs for an eleven-element array.

    The accurate delay model was calculated using the Goddard Space
Flight Center (GSFC) CALC software package version 9.1\footnote{
www.sgl.crestech.ca/IVS-Analysis/software\_tools\\/calc\_solve/datafiles.htm},
which uses the most recent parameters associated with the earth
rotation and orientation, nutation, the terrestrial reference frame,
and the radio source positions defined on the celestial reference
frame \citep{ma98}.  We have slightly modified the CALC package to
incorporate the gravitational deflection from all significant solar
system bodies including the retardation term, as given by Eq.\ (1).
We have assumed that the PPN parameter $\gamma=1$, because it has been
measured to 0.1\% accuracy \citep{leb95,eub97}.  We could not improve
this value in our experiment, and its uncertainty produces a deflection
offset $\le 1~\mu$arcsec, much less than the experimental accuracy
that we achieved.  Parameters from which telescope delay variability
could be determined were monitored during the observations
(egs. ground weather conditions), and ionospheric data was available
from GSFC data archives collected by Global Positioning Satellites
(GPS).

    The phase part of the complex visibility function (called simply
the phase) for each source $a$, can be written as
\begin{eqnarray}
   \phi^a_{l,m}(t) = \frac{\nu}{c}\Bigl[
        {\bf B}_{l,m}{\bf \cdot} {\triangle{\bf K}}^a(t) +~~~~~~~~~~~ \nonumber \\
        {\triangle}{\bf B}_{l,m}(t)\cdot{\bf K}^a +
        {\triangle}C_{l,m}(t)+{\triangle}A^a_{l,m}(t) \Bigr] +
        \psi^a_{l,m}(t)
\end{eqnarray}
\noindent
where ${\bf B}_{l,m}$ is the model separation between telescope $l$
and $m$, $\triangle{\bf B}_{l,m}(t)$ is the separation error; ${\bf
K}^a$ is the model position of the $a$th source, $\triangle{\bf
K}^a(t)$ is the position offset error for the $ath$ source,
{$\triangle$}C$_{l,m}(t)$ is the residual clock delay between
telescopes, and $\triangle A^a_{l,m}(t)$ is the residual
tropospheric/ionospheric delay in the direction to the $ath$ source.
The structure of the source is given by the phase part of the complex
spatial coherence function, $\psi^a_{l,m}(t)$.  For this experiment, $a$
has three values: $a=0$ (J0842+1835), $a=1$ (J0839+1802), $a=2$
(J0854+2006).  The occasional lobe ambiguities (arbitrary turns of
phase) of the phase measurements were easily determined because of the
accuracy of the delay model.

\subsection {The Source Structure}
   The accuracy of the experiment depends in part on the stability of
the three sources between September 4 to 12.  In terms of variables in
Eq.\ (6), the intrinsic change of position, $\triangle{\bf K}^a(t)$,
of each source, and the visibility structure phases variations,
$\psi^a_{l,m}(t)$, must be small.  The structures for each source on
each day were determined from their respective complex visibility
function using self-calibration techniques for radio imaging
\citep{wal95}.  This imaging/deconvolution iteration process does {\it
not} determine the accurate positions of each source, only its shape.
The derived images of the three quasars during the experiment are
shown in Fig.\ 1.  All three sources show the typical structure
associated with most luminous radio sources, a bright component (core)
at one end of the structure containing an appreciable part of the
emission, and more extended emission, often with a secondary bright
component.  There were no apparent changes in the intensity and shape
of the sources, except for small changes in J0854+2006 at the 1\%
contour level.  These properties are in agreement with previous
observations of J0842 (sources names will be abbreviated) and J0839
which show little long term structure changes \citep{fey00,bea02},
whereas J0854 is a known variable source.

    The observation dates are symmetric with respect to September 8.
Hence, even if there are long-term small structure changes over the
experiment, the deviation of the average structure phase of each
source, based on observations averaged on September 4,7,9 and 12,
should be equal to first order to that on September 8 (this averaging
is discussed below).  Nevertheless, the change in the measured phase
associated with the structure difference in J0854 between September 4
and 12 is less than 0.003 turn for any baseline, and this
corresponds to an effective positional error of less than
$3~\mu$arcsec, even without the above averaging which was incorporated
in the reductions.  Another indication of very little change over the
7-day duration of the observations is that the separation of the radio
core and the secondary component for all sources was stable to $\le
20~\mu$arcsec, an accuracy consistent with the sensitivity of the
observations and the somewhat diffuse nature of the secondary peak.

   The large distance to the radio sources guarantees that any proper
motion will be much less than one microarcsec over the week period of
the experiment.  However, some compact components ($<50~\mu$arcsec) in
radio sources vary slightly in intensity over hour time-scales because
of galactic scattering at a distance of $\approx 300$ pc
\citep{big03}.  It is possible that such scattering also irregularly
moves the apparent position of the component.  However, we detected no
short-term intensity variations in any source greater than 4\% over
periods of a few hours, and concluded that the compact components are
not affected by interstellar scattering.  However, we emphasize that
such movement of the source position, if it existed, could be detected
in this experiment as larger than expected residual phases with time.

\subsection {Removing the Temporal and Spatial Phase Errors}
    With the removal of the source structure terms (we used the
average source structure over the five observing days), the phase in
Eq.\ (6) becomes circular among the telescopes; that is
$\phi_{l,m}=\phi_{l,n}+\phi_{n,m}$, apart from stochastic noise.
Therefore, there are only 10 independent phases for an array of
11-telescopes.  We can thus choose a reference telescope, $M$, and
write Eq.\ (6) for each source $a$ and telescope $l$, at time $t_a$ as
\begin{eqnarray}
   \phi^a_{l,M}(t_a) = f_{l,M}(t_a) + \nonumber \\ 
                {\bf g}_{l,M}(t_a,{\bf K}^a) + 
               \triangle{\bf K}^a(t_a){\bf\cdot} {\bf B}_{l,M}
\end{eqnarray}
where $f_{l,M}(t_a)$ is a function of time, which is dominated by the
clock drifts and other instrumental variations at each telescope, and
is independent of the source position.  The term ${\bf
g}_{l,M}(t_a,\bf{K}^a)$ describes the temporal and spatial properties
of the delay screen above each telescope near the position of the
three sources.  The last term in Eq.\ (7) cannot be written as part of
the first two terms since it depends on an unknown, but small, offset
for each source from the assumed value.  We used Mauna Kea (MK), HI as
the reference telescope for most of the observations since this
telescope is at a high site with generally good atmospheric
conditions.  For the first three hours of each day when the source
elevation at MK was $<20^\circ$, we used the Los Alamos, NM telescope
as the reference telescope.

  The properties of the several components of delay which contribute
to the time variable phase screen over each telescope are well-known.
The phase error associated with the uncertain location of a telescope
${\bf \triangle B}_{l,M}$, which includes any deviation of the earth
rotation and orientation and nutation terms from those used in
correlator CALC Model, are smooth functions of time and direction in
the sky.  The component produced by the changing refractivity of the
troposphere and ionosphere toward the sources from relatively large
{\it clouds} and diurnal changes persists for tens of minutes and are
generally larger than $5^\circ$ in angular size.  On the other hand,
small tropospheric clouds and ionospheric events produce
quasi-stochastic changes above each telescope.  In their extreme
behavior, these rapidly changing delays have time scales less than a
few minutes and degrees, making astrometric VLBI observations useless.

We can expand the first two terms in Eq.\ (7) into a Taylor series
around $(t_a-t_0)$ and $({\bf K}^a-{\bf K}^0)$.  Since J0842 and the
two calibrators lie nearly linearly in the sky, the angular part of
the expansion is only needed in the direction of the source alignment
in the sky\footnote{Three calibrator sources are needed in the general
case.}.  The first order terms can be estimated by a sum of the
observed calibrator phases, weighted by their separation from J0842 in
the sky, and in time.  Thus, the calibrated phase for J0842,
$\Phi(t_0)_{l,M}$ at its observation time, $t_0$, for any day for
telescope $l$ becomes
\begin{eqnarray}
\Phi(t_0)_{l,M}=\phi^0(t_0)_{l,M} - \left(0.80\phi^1(t_0)_{l,M}
     +0.20\phi^2(t_0)_{l,M}\right)\\ \nonumber
     \approx {\bf B}_{l,M}{\bf\cdot}\Big(\triangle{\bf  K}^0(t_0)-0.80\triangle
          {\bf K}^1(t_0)-0.20\triangle {\bf K}^2(t_0)\Big)
\end{eqnarray}
\noindent where $\phi^1(t_0)$ and $\phi^2(t_0)$ are the phases of
sources 1 or 2, each measured at times $t_1$ and $t_2$, but
linearly-interpolated to time $t_0$.  In other words, the algorithm in
Eq.\ (8) defines how the measured phases for J0839 and J0854 are
combined to remove, to first order, the temporal and angular phase
change in the sky at the position of J0842 at the time of its
observation at $t_0$.  The calibrated phase of J0842, thus, depends
explicitly on the change of position of the three radio sources with
time.  The short time-scale and small angular-scale fluctuations are
not removed, but these are nearly stochastic and average out over
longer periods of time.  They add to the uncertainty of the results,
but do not contribute to any systematic biases.  Second order terms,
which are not removed from the interpolation used in Eq.\ (8), have
been minimized by using precise apriori parameters available from the
CALC package and available tropospheric and ionospheric data
associated with each telescope from the GPS data base.

    Examples of the effectiveness of this calibration process are
shown in Fig.\ 2 for the Owens Valley, CA to Mauna Kea, HI baseline on
September 9.  The observed phases, $\phi^a(t)$, (top plot) for the
three sources follow each other over the day, but are separated at any
time by an amount which is varying, caused by a phase wedge in the
atmosphere.  The displacement of phase between J0854 and J0842 is
larger and in the opposite sense than that from J0839 and J0842, as
expected from linear phase wedge.  The calibrated phase $\Phi$ of
J0842, (bottom) is relatively smooth and near zero phase.  The
relatively small scatter is caused by fluctuations of phase shorter in
temporal scale than about five minutes, or smaller in angular scale
than five degrees.  The signal to noise inherent in the observations
cause only a small part of the scatter.

     There are periods of time when the phase stability shown by all
three sources noticeably deteriorates, and the calibrated phase of
J0842 becomes large.  These periods often occur when the source is at
a low elevation, typically less than $20^\circ$, when large and
non-linear tropospheric and ionospheric phase are more likely.
Occasionally at high elevations during periods of rainy and windy
conditions, VLBI observations are useless.  These periods of obvious
poor phase stability were removed from further consideration, and
consisted of 15\% of the original data.  On the average the phase
stability at sites with dry conditions was more stable than at those
with humid conditions.

   The calibrated phases, $\Phi_{l,M}(t)$, were then averaged for each
day over a one hour period, with an estimated rms error determined from
the scatter, which is about 0.02 turns per point.  In Fig.\ 3, we show
these averaged calibrated phases for two telescopes.  The calibrated
phases repeat extremely well from day to day to an rms scatter of about
0.02 turns.  Even a slight negative offset at GST$\approx 15$ hr
repeats, and is produced by a $\sim 50~\mu$arcsec offset of the source
position from that assumed in the model, as given in Eq.\ (8).

\subsection {The Deflection on September 8}

The difference between the calibrated phase on September 8, and the
average of the phases from the other days, is shown in Fig.\ 4 for the
same two telescopes as in Fig.\ 3.  Assuming that the position of all
three sources remain fixed with respect to the apriori model, which
includes the radial and retarded gravitational bending predicted by
GR, the phases in Fig.\ 4 should scatter around zero phase,
corresponding to $\delta=0$.  The curve given by $\delta=-1$ shows the
expected calibrated phase on September 8 if the retarded deflection
term were zero.  Fig.\ 4 already demonstrates that $\delta\approx 0$,
and that the experiment has clearly detected the retarded component of
the light deflection.

     An analysis of the effect of the jovian magnetosphere, has been
described fully \citep{kop02}, and is summarized here.  The Galileo
spacecraft provided an estimate of the average electron plasma
density, $N_0$, near the jovian surface.  We assumed a
spherically-shaped magnetosphere with a plasma density which decreases
as $N_0(R_J/r)^{2+A}$ (where $R_J=7.1\times 10^7$ m is the mean radius
of Jupiter and A is an exponent reflecting different magnetospheric
models), and integrated the estimated plasma delay at 8.45 GHz
along the propagation path of J0842 at closest approach ($3.7'$ or
$13R_J)$.  We obtain magnetospheric deflection estimates of 17.5, 1.0,
0.02~$\mu$arcsec for $A=$0.0, 1.0, 2.0, respectively.  Based on the
analogy with radial distribution of the solar corona, which suggests
$A\approx 0.33$, we have determined the position shift associated with
this jovian magnetosphere model (it is in the opposite direction of
the gravitational radial deflection of light), and we have shown its
estimated phase contribution by the dashed line in Fig.\ 4.  This
estimate is probably substantially larger than the actual value on
this day.

    There are two ways in which we can combine the data in order to
determine the mean gravitational deflection of J0842.  First, we
averaged the data for each telescope, and determined $\delta$
associated with each telescope, and these results are shown in Table
1, listed in decreasing order of accuracy.  We find that the most
accurate determinations are associated with telescopes in the
south-west USA, about 5000 km from MK, because they are located at
places with a relatively stable and dry atmosphere.  Telescopes in more
humid locations give poorer results.  The Effelsberg telescope,
however, was critical in the determination of the accurate structure
of the radio sources.  All telescope determinations are consistent
with $\delta=0$ when their estimated errors are considered.

    Another method of display is to produce an image from the
calibrated phase data for all baselines, which are measured during any
one-hour period.  The location of the peak of the image is the
position of J0842, and these positions are shown in Fig.\ 5.  The
east/west position of the source is more accurately determined than
the north/south position since the array spans mostly in the east/west
direction.  The east/west positions clearly show that the retarded
component of the light deflection has been detected, and is consistent
with GR.  The red line shows the estimate of a reasonable jovian
magnetospheric refraction, and is much smaller than the retarded
deflection, particularly in the east/west direction.

    The weighted average value of the data for the observations over
September 8, regardless of how the data are averaged (by telescope or
by time), is
$$\delta=-0.02\pm 0.19~~~~~~~~\hbox {1-$\sigma${~error}}$$
\noindent 
Hence, the measured retarded deflection associated with the retarded
position of Jupiter given by Eqs.\ (1)-(3), is in good agreement with
the prediction of GR.  Our interpretation of $\delta$ sets a bound of
$c_g=(1.06\pm 0.21)c$ (see \S 4.b)
\section {Discussion}

\subsection{The Measurement Results}

The measurement of the retarded deflection component in Eq.\ (1) is an
accurate experimental confirmation of General Relativity in the solar
system, using the motion of Jupiter.  To determine $\delta$ with
sufficient accuracy, improvements in the technique of phase-reference
VLBI were developed.

    The use of two calibrators, on either side of J0842, improved the
positional accuracy of this phase-referencing VLBI experiment by a
factor of more than three compared with previous observations of this
type.  Because of the relative simplicity of the experimental concept
as a classical deflection-type experiment, the uniform observational
procedure over the entire experiment, the data redundancy provided by
the number of telescopes and the number of days, we are convinced that
the result and error estimate are accurate and demonstrated from the
Fig.\ 4 and Fig.\ 5 which display a significant part of the data.  The
refraction from the jovian magnetosphere is at least as small as we
have estimated \citep{kop02}.  Any position jitter caused by
interstellar scattering is not significant since the scatter in Fig.\
5 around the expected retarded position is no larger than that
expected from the errors derived from the scatter in the observations
among all of the days.

    The two-calibrator technique, developed in this experiment to
remove the delay variations of the tropospherically-induced phase
wedge above the telescopes, can be used for determinations of the
proper motion and parallax of radio sources.  With the relative
positional sensitivity of $10~\mu$arcsec, the trigonometric parallax
of stars and pulsars as far as 10 kpc can be determined; spacecraft
can be tracked with an accuracy of 30 m near Jupiter; and the proper
motion of radio sources in M31 can be determined to 5 km/sec
($1~\mu$arcsec/yr) over ten years.  However, the significant structure
changes in many calibrator sources over periods longer than a few
months may dominate the error budget at angular levels of
$>50~\mu$arcsec for the long-term astrometric projects.  Expected
improvements in the accuracy of the celestial frame over the next
decade may permit the determination of calibrator core positions and
changes needed to reach $10~\mu$arcsec accuracy \citep{ma98, jac02}

\subsection{The Measurement Interpretation}

Our interpretation of this novel deflection experiment comes directly
from the formulation of \cite{kop01,kop03} where the retarded
deflection term (and the availability of the September 8, 2002 Jupiter
experiment) was initially proposed.  The interpretation is that the
position of Jupiter at the VLBI delay given by Eq.\ (1) must be taken
at the retarded time with respect to the time of observation, as
defined by the solution of the gravity null cone Eqs.\ (2) and (3).
This retardation is due to the finite speed of gravity and can be
observed.  The measured retarded deflection, as expressed by the
parameter $\delta$, is in agreement with GR to 20\% accuracy, and that
the formulation, summarized by Eqs.\ (1)-(3), is confirmed to {\it the
first order in v/c}.  If $\delta\ne 0$, then the constant $c_g$ in
the wave operator of the Einstein equations is not equal to the speed
of light numerically, and GR (the gravity null cone) would not be
invariant with respect to the Lorentz transformation.

Another interpretation has been proposed by \cite{wil03} who made use
of a matching technique that was first employed for solving the same
problem by \cite{kli92}.  This formulation {\it assumes} that the
position of Jupiter in Eq.\ (1) must be taken at the time of the
closest approach of light to Jupiter.  Will's formulation is not
Lorentz-invariant but approximates the Lorentz-invariant formulation
of the time delay up to linearized terms of order $v/c$.

Both formulations\footnote{We suggest that the reader compare the
treatment of the September 8, 2002 experiment by \cite{kop01,kop03},
by \cite{wil03}, by \cite{asa02}, for a fuller understanding of the different
interpretations and mathematical techniques involved in derivation of
the basic equations.  Papers by \cite{kop99,kli92,kli03,car00} may be
valuable as well.} give identical predictions for the deflection of
light by a moving body to first order in $(v/c)$.  However, the
physical meaning attached to the measurement is quite different.  The
\cite{kop01} interpretation is related to the Lorentz-invariant
definition of the gravity null cone and the finite speed of
propagation of gravity as predicted by the Li\'enard-Wiechert solution
of Einstein's equations.  \cite{kop03} also associates the $v/c$
effects with the gravitomagnetic dragging of the light ray caused by
the translational motion of Jupiter with respect to the barycenter of
the solar system.  The \cite{wil03} and \cite{asa02} interpretation is
that the $v/c$ term is related to the aberration of light that diverges
from our interpretation at terms of order $v^2/c^2$.  Our
measurement, unfortunately, is not sufficiently accurate to detect
terms of order $v^2/c^2$ which can distinguish between the existing
interpretations experimentally.  This could be a challenge for the
next generation of space VLBI or other astrometric space missions like
SIM or GAIA.  Relativistic deflection terms of order $v^2/c^2$ for
light grazing Jupiter's limb, may reach a magnitude of 1 $\mu$arcsec
and can be, in principle, observable.

    What is clear is that if this experiment had measured $\delta$
significantly different from zero, then a problem would exist with the
current formulation of General Relativity and, as such, any
interpretation of the result would be ambiguous and depend on the
theoretical generalization of the Einstein equations that was chosen.
However, this measurement of the deflection of light by Jupiter is in
good agreement with General Relativity and demonstrates experimentally
that the Einstein Equations are Lorentz-invariant and the retarded
position of Jupiter in Eq.\ (1) should be used in the determination of
the deflection of light in accordance with Eqs.\ (2) and (3).

\acknowledgments

The National Radio Astronomy Observatory is a facility of the National
Science Foundation, operated under cooperative agreement by Associated
Universities, Inc.  We thank the support from the Department of
Physics and Astronomy, College of Arts and Science, and the Research
Council of the University of Missouri-Columbia.  The Max-Planck-Institut
f\"{u}r Radioastronomie operates the Effelsberg 100-m Radio Telescope.
We thank the VLBA staff for their help and support in obtaining and
correlating the data. Support from the Eppley Foundation for Research
(award \#002672) and the Technical Research Council of Turkey (TUBITAK)
is greatly appreciated.  SK acknowledges the Department of Physics of
the Istanbul Technical University for their hospitality.  We thank
Norbert Bartel and two anonymous referees for their comments and
suggestions.

\clearpage

\begin{deluxetable}{lcc}
\tablewidth{0pt}
\tablecolumns{3}
\tablecaption{Telescope Solutions for $\delta$}
\tabletypesize{\normalsize}
\tablehead {
   \colhead {Telescope}  &
   \colhead {Baseline to MK}  & 
   \colhead {$\delta$ } \\
   \colhead { }  &
   \colhead {(km)} \\
}
\startdata
Hancock, NH          &  7500  & $- 0.02\pm 0.25$ \\
Los Alamos, NM       &  4970  & $- 0.20\pm 0.34$ \\
Kitt Peak, AZ        &  4490  & $+ 0.52\pm 0.38$ \\
Owens Valley, CA     &  4015  & $- 0.45\pm 0.42$ \\
Pie Town, NM         &  4800  & $+ 0.30\pm 0.44$ \\
Fort Davis, TX       &  5130  & $- 0.09\pm 0.52$ \\
North Liberty, IA    &  6160  & $- 0.37\pm 0.53$ \\
Brewster, WA         &  4400  & $+ 0.14\pm 0.62$ \\
Saint Croix, VI      &  8610  & $- 0.82\pm 0.85$ \\
Effelsberg, Germany  & 10300  & $+ 1.94\pm 1.60$ \\
${\bf <\delta>}$              &        &  ${\bf- 0.02\pm 0.19}$ \\
\enddata
\end{deluxetable}

\end{document}